\documentclass[12pt]{article}
\begin{document}

\addtolength{\baselineskip}{0.5\baselineskip}

\title{\textbf{Hierarchy Bloch Equations for the Reduced Statistical Density Operators
in Canonical and Grand Canonical Ensembles}}
\author{Liqiang Wei\\
Institute for Theoretical Atomic, Molecular and Optical Physics\\
Harvard University, Cambridge, MA 02318\\
\\
Chiachung Sun\\
Institute of Theoretical Chemistry, Jilin University\\
Changchun, Jilin 130023 P. R. China}

\maketitle

\begin{abstract}
\vspace{0.05in}
 Starting from Bloch equation for a canonical ensemble, we deduce
 a set of hierarchy equations for the reduced statistical density
 operator for an identical many-body system with two-body
 interaction. They provide a law according to which the reduced
 density operator varies in temperature. By definition of the
 reduced density operator in Fock space for a grand canonical
 ensemble, we also obtain the analogous Bloch equation and the
 corresponding hierarchy reduced equations for the identical
 interacting many-body system. We discuss their possible
 solutions and applications.
\\
\\
$\underline{PACS}$ 05.30.-d; 05.30.Fk; 05.30.Jp; 31.90.+s;
71.10.-w
\end{abstract}

\vspace{0.35in}
\section{Introduction}

The fundamental problem in quantum statistical mechanics is to
determine the density matrix of a statistical ensemble. Once the
density matrix is known, the thermodynamic properties can be
calculated from the corresponding microscopic mechanical
observables [1-4]. The basic equation of motion for the
statistical density matrix is von Neumann-Liouville equation. For
an equilibrium statistical ensemble, its solution takes the
following special form,
\begin{equation}
\rho = \sum_{i} \rho_{i}|\psi_{i}><\psi_{i}|,
\end{equation}
where $\rho_{i}$ is the thermal probability distribution, and
$|\psi_{i}>$ is the energy eigenstate of the system. The summation
over $i$ is for all the eigenstates consistent with symmetry
constraints. For a canonical ensemble, for example, its relative
thermal probability distribution and density operator are as
follows,
\begin{equation}
  P_{i} = \exp (-\beta E_{i}),
\end{equation}
and
\begin{equation}
   D^{N} = \sum_{i} \exp(-\beta
   E_{i})|\psi_{i}><\psi_{i}|=\exp(-\beta H_{N}),
\end{equation}
where $\beta$ is the reciprocal of the product of Boltzmann's
constant $k_{b}$ and the absolute temperature $T$, and $H_{N}$ is
the Hamiltonian of an $N$-body system. The trace of Eq. (3) gives
the partition function of the canonical ensemble,
\begin{equation}
  Tr(D^{N}) = \sum_{i}\exp(-\beta E_{i}) = Z(\beta, V, N).
\end{equation}

From their definition in Eqs. (1) or (4), we observe that, to
determine the density matrix or the partition function, there are
generally two steps involved. In the first step, we have to solve
an energy eigenequation for the interacting many-body system to
get its energy eigenstates and eigenvalues. In the second one, we
need to sum over all the microscopic states. Obviously, these are
both a very difficult or prohibitive task for the system
considered, especially when the quantum effects are important and
can not be neglected. Nevertheless, some other approaches or
approximations have been developed, including the one called the
equation of motion method, to address these issues [5,6].
Furthermore, for an identical interacting many-body system, it is
known that its full density matrix conveys much more information
than is necessary, and its reduced density matrix suffices to
describe the states of a whole system. In analogous to that for
the full density matrix, we attempt here to develop similar
equations for the reduced density matrix so that they provide a
kind of laws that the reduced density matrix obeys, or from which,
the reduced density matrix can be directly determined.

This paper is organized as follows. In the next section, we derive
the reduced Bloch equations for the reduced statistical density
operators for a canonical ensemble. They are the equations of
motion according to which the reduced density matrix varies in
temperature. In Section 3, we define the reduced density matrix
for a grand canonical ensemble, and obtain the analogous Bloch
equation. We also extend the ideas developed in Section 2 on the
reduced Bloch equation from the canonical ensemble to the case of
a grand canonical ensemble. In Section 4, we summarize our results
and discuss the possible routes for solving the equations as well
as their applications.

\vspace{0.35in}
\section{Bloch Equation for the Reduced Statistical Density
Operators in Canonical Ensemble}

Consider a canonical ensemble for an interacting $N$-body system
whose density matrix is shown in Eq. (3). Differentiation of both
sides of Eqs. (2) or (3) with respect to $\beta$ leads immediately
to the differential equation,
\begin{equation}
-\frac{\partial}{\partial\beta} D^{N} = H_{N} D^{N},
\end{equation}
which is called the Bloch equation [5,6]. It provides a law
according to which the canonical density operator varies in
temperature. It is equivalent to the definition (3). Nevertheless,
it offers an alternative route to determine the density matrix. If
the Hamiltonian of the system is known, we can directly solve the
exact or approximate density matrix without recourse to the two
procedures as we outline in the Introduction Section. A noticeable
feature of Eq. (5) is that after its substitution of $\beta$ by
$it/\hbar$, it yields an equation that is identical in form with
the time-dependent Schr$\ddot{o}$dinger equation. Therefore, some
developed techniques for solving the time-dependent
Schr$\ddot{o}$dinger equation can be borrowed to obtain the
solutions for the Bloch equation (5).

In practical applications, moreover, the interactions among the
identical particles such as electrons in atoms, molecules, and
solids are often two-body forces. The $N$th-order density matrix
contains much more information than is required. In order to
calculate the thermodynamic properties of the system, we need at
most the second-order reduced density matrix and not the full
$N$th-order one. That is to say, the $p$th-order reduced density
matrix can describe the $N (>p\ge 2)$ particle states. For this
purpose, we try to find a method for directly solving the
$p$th-order reduced density matrix instead of the one for a full
$N$th-order density matrix.

The $p$th-order reduced density matrix of an $N$-particle system
is generally defined by [7-11]
\begin{equation}
D^{p} = L^{p}_{N} (D^{N}),
\end{equation}
where $L^{p}_{N}$ is the contraction operator [12,13]. Its trace
gives the partition function defined before
\begin{equation}
  Tr(D^{p}) = Z(\beta, V, N).
\end{equation}
Assume that the Hamiltonian of the system considered is made up of
two contributions,
\begin{equation}
  H_{N} = \sum_{i=1}^{N} h(i) + \sum_{i<j}^{N} g(i,j),
\end{equation}
where $h(i)$ and $g(i,j)$ are one- and two-particle operators,
respectively. Rewrite the $H_{N}$ into
\begin{equation}
H_{N}=H^{p}_{1}+\sum_{j=p+1}^{N}h(i)+\sum_{i=1}^{p}\sum_{j=p+1}^{N}g(i,j)
+\sum_{i<j\\(i\ge p+1)}^{N} g(i,j),
\end{equation}
where
\begin{equation}
  H_{1}^{p} = \sum_{i=1}^{p} h(i) + \sum_{i<j}^{p} g(i,j).
\end{equation}
Then substituting the Hamiltonian Eq. (9) into the Bloch equation
(5) and applying on both sides the operator $L^{p}_{N}$, we get
the following equations for the reduced density operators,
\begin{eqnarray}
\nonumber
-\frac{\partial}{\partial\beta}D^{p}&=&H^{p}_{1}D^{p}+(N-p)L^{p}_{p+1}
\left[h(p+1)D^{p+1}\right]+(N-p)L_{p+1}^{p}\left[\sum_{i=1}^{p}g(i,p+1)D^{p+1}\right]+\\
&&+\left(\begin{array}{c} N-p\\2\end{array}\right)
L_{p+2}^{p}\left[g(p+1,p+2)D^{p+2}\right].
\end{eqnarray}
We call them the reduced Bloch equations for the canonical
ensemble. They relates the $D^{p}$, $D^{p+1}$, and $D^{p+2}$ in a
certain way, and can be seen as a set of hierarchy. From the
general definition for the reduced density matrix (6), they can
also be regarded as an independent equation for $D^{p+2}$. Similar
procedures have lead to the hierarchy equations for the reduced
density matrix of the pure states before from the
Schr$\ddot{o}$dinger equation in a density matrix form [14-17]. It
follows from the idea in deriving the Bogoliubov-Born-Green-Yvon
(BBGKY) hierarchy in classical statistical mechanics [4].

\vspace{0.35in}
\section{Bloch Equation for the Reduced Statistical Density
Operators in Grand Canonical Ensemble}

To provide more avenues for studying the behaviors of statistical
density matrices, the above description can also be extended to
the case of a grand canonical ensemble. In $N$-particle Hilbert
space $V^{N}$, the relative grand canonical distribution reads
\begin{equation}
  P_{i}(N) = \exp[-\beta(E_{i}-\mu N)],
\end{equation}
and the density matrix is
\begin{eqnarray}
\nonumber D_{G}(N)&=& \exp[-\beta(H-\mu N)],\\
  &=& \exp(-\beta \bar{H}),
\end{eqnarray}
where
\begin{equation}
  \bar{H} = H - \mu N,
\end{equation}
is called the grand Hamiltonian on $V^{N}$. Again, upon
differentiating both sides of Eq. (13) with respect to $\beta$, we
have
\begin{equation}
-\frac{\partial}{\partial\beta} D_{G}(N) = \bar{H} D_{G}(N),
\end{equation}
which has the same form as Eq. (5). We call it the Bloch equation
for the grand canonical ensemble on $N$-particle Hilbert space
$V^{N}$.

For the grand canonical ensemble, the number of particles $N$
fluctuates. Its density matrix should be defined in the entire
Fock space: $F = \sum_{N}\oplus V^{N}$. That is, it should be
written as the direct sum of every density matrix $D_{G}(N)$
belonging to the $N$-particle space $V^{N}$,
\begin{equation}
   D_{G} = \sum_{N=0}^{\infty} \oplus D_{G}(N),
\end{equation}
and its trace in Fock space is the grand partition function,
\begin{equation}
  Tr(D_{G}) = \sum_{N=0}^{\infty}\sum_{i}\exp[-\beta(E_{i}-\mu
  N)]=\Xi(\beta, \mu, V).
\end{equation}
For the same consideration, we define the corresponding
$p$th-order reduced density matrix as follows,
\begin{equation}
 D^{p}_{G} =\sum_{N=p}^{\infty}\oplus
 \left(\begin{array}{c}N\\p\end{array}\right)L_{N}^{p}[D_{G}(N)],
\end{equation}
where the binomial coefficient
$\left(\begin{array}{c}N\\p\end{array}\right)$ is added to reflect
the symmetric property of the density matrix in terms of the
permutation of particles. Obviously,
\begin{equation}
  D_{G}^{0} = \Xi(\beta, \mu, V),
\end{equation}
and
\begin{equation}
  Tr(D_{G}^{p}) =
  \left<\left(\begin{array}{c}N\\p\end{array}\right)\right>
  D^{0}_{G}.
\end{equation}
Assume Hamiltonian $H$ in $\bar{H}$ takes the same form as in Eq.
(8). Rewrite $\bar{H}$ into
\begin{equation}
\bar{H}=\bar{H}^{p}_{1}+\sum_{j=p+1}^{N}\bar{h}(i)+\sum_{i=1}^{p}\sum_{j=p+1}^{N}g(i,j)
+\sum_{i<j\\(i\ge p+1)}^{N} g(i,j),
\end{equation}
where
\begin{equation}
 \bar{h}(i) = h(i)-\mu,
\end{equation}
and
\begin{equation}
  \bar{H}_{1}^{p} = \sum_{i=1}^{p}\bar{h}(i) + \sum_{i<j}^{p} g(i,j).
\end{equation}
Substituting the Eq. (21) into Eq. (15) and applying on both sides
the operator $\sum_{N=p}^{\infty}\oplus
\left(\begin{array}{c}N\\p\end{array}\right) L^{p}_{N}$, we
finally obtain the equations for the reduced density matrix
\begin{eqnarray}
\nonumber
-\frac{\partial}{\partial\beta}D^{p}_{G}&=&\bar{H}^{p}_{1}D^{p}_{G}+(p+1)L^{p}_{p+1}
\left[\bar{h}(p+1)D^{p+1}_{G}\right]+
(p+1)L_{p+1}^{p}\left[\sum_{i=1}^{p}g(i,p+1)D^{p+1}_{G}\right]+\\
&&+\left(\begin{array}{c} p+2\\2\end{array}\right)
L_{p+2}^{p}\left[g(p+1,p+2)D^{p+2}_{G}\right],
\end{eqnarray}
which are the reduced Bloch equation on Fock space for the grand
canonical ensemble. They also represents a law accoding to which
the grand reduced density matrix changes in temperature. Similar
notes can also be made as in the case for the canonical ensemble.

\vspace{0.35in}
\section{Summary and Discussions}

In this paper, within the framework of equilibrium quantum
statistical mechanics, we have derived the hierarchy equations of
the density operators for both canonical and grand canonical
ensembles. They are the equations of motion in terms of the
variation of temperature that the reduced density operator
satisfies. They provide the route for a direct determination of
the reduced density operator for the statistical ensembles.

In the first place, the Eqs. (11) or (24) can be solved in an
approximate way. Since they are a set of hierarchy equations, we
have to relate the $D^{p+1}$ and the $D^{p}$ in a certain way to
decouple them. The most obvious choice is to take an orbital
approximation. For an identical fermion system, this means to
express a $p$th-order reduced density matrix as a Grassmann
product of $p$ first-order reduced density matrices as follows
\begin{eqnarray}
\nonumber D^{p+1}&=&D^{p}\wedge D^{1}/D^{0} \\
&=& \underbrace{D^{1}\wedge D^{1} \wedge ... \wedge
D^{1}}_{p+1}/(D^{0})^{p}.
\end{eqnarray}
Similar scheme has been applied for solving the reduced
Schr$\ddot{o}$dinger equation for the reduced density matrix to
obtain the usual Hartree-Fock equation [14-17]. We expect that it
will lead to more general equation beyond that for zero
temperature when it is applied to the Eq. (24) [21]. Consequently,
this will provide a possibility for studying the microscopic
structure of high temperature species or macroscopic molecules at
different temperatures [19-22]. In addition, it will serve as a
staring point to develop more general schemes for studying such
important issues as the interplay between the microscopic
structure and macroscopic thermodynamic properties [18]. This also
includes the pressure effects on the electronic structure for
solids [23]. Furthermore, Eqs. (11) or (24) can also be solved
exactly for some model systems. One example is to consider a
pairing Hamiltonian, and solve the equations in an occupation
number representation [25-27]. The pairing Hamiltonian is a useful
model, for example, in the study of the nuclear structure and the
microscopic states of superconductors. We hope that we can obtain
its exact reduced density matrices with different orders and
therefore investigate the corresponding thermodynamic properties
[28]. Work along these two lines can be carried out.


\vspace{0.45in}

\begin{thebibliography}{99}
\vspace{0.15in}

\bibitem{tolman} R. C. Tolman, The Principles of Statistical
Mechanics (Oxford University Press, 1938).
\bibitem{feynman} R. P. Feynman, Statistical Mechanics (W. A.
Benjamine, Reading, Mass., 1972).
\bibitem{chandler} D. Chandler, Introduction to Modern Statistical
Mechanics (Oxford University, 1987).
\bibitem{huang} K. Huang, Statistical Mechanics (John Wiley \&
Sons, 2nd Edition, 1987).
\bibitem{bloch} F. Bloch, Zeits. f. Physick 74, 295 (1932).
\bibitem{kirkwood} J. G. Kirkwood, Phys. Rev. 44, 31 (1933).
\bibitem{husimi} K. Husimi, Proc. Phys. Math. Soc. Japan, Ser. 3
32, 264 (1940).
\bibitem{lowdin1} P. O. L$\ddot{o}$wdin, Phys. Rev. 97, 1474
(1955).
\bibitem{fano} U. Fano, Rev. Mod. Phys. 29, 74 (1957).
\bibitem{mcweeny} R. McWeeny, Rev. Mod. Phys. 32, 335 (1960).
\bibitem{davidson} E. R. Davidson, Reduced Density Matrices in
Quantum Chemistry (New York: Academic Press, 1976).
\bibitem{kummer} K. Kummer, J. Math. Phys. 8, 2063 (1967).
\bibitem{harriman1} J. E. Harriman, Phys. Rev. A 17, 1257 (1978).
\bibitem{cohen} L. Cohen and C. Frishberg, Phys. Rev. A 13, 927
(1976).
\bibitem{nakatsuji} H. Nakatsuji, Phys. Rev. A 14, 41 (1976).
\bibitem{schlosser} H. Schlosser, Phys. Rev. A 15, 1349 (1977).
\bibitem{harriman2} J. E. Harriman, Phys. Rev. A 19, 1893 (1979).
\bibitem{lowdin2} P. O. L$\ddot{o}$wdin, Intern. J. Quantum Chem.
Vol xxix, 1651 (1986).
\bibitem{klemperer} W. Klemperer, Ann. Rev. Phys. Chem. 46, 1
(1995).
\bibitem{friend} K. Pichler, D. A. Halliday, D. D. C. Bradley, P.
L. Burn, R. H. Friend, and A. B. Holmes, J. Phys.: Condens. Matter
5, 7155 (1993)
\bibitem{lin} J. Yu, M. Hayashi, S. H. Lin, K.-K. Liang, J. H. Hsu, W.
S. Fann, C.-I. Chao, K.-R. Chuang, S.-A. Chen, Synth. Met. 82, 159
(1996).
\bibitem{deleuze1} S. P. Kwasniewski, J. P. Francois, and M. S.
Deleuze, J. Phys. Chem A 107, 5168 (2003).
\bibitem{drickermer} H. G. Drickamer and C. W. Franck, Electronic
Transitions and the High Pressure Chemistry and Physics of Solids
(London, Chapman and Hall, New York, 1973).
\bibitem{wei1} L. Wei and C. C. Sun, submitted (2003); cond-mat/0306307.
\bibitem{kisslinger} L. S. Kisslinger and R. A. Sorensen, Mat.
Fys. Medd. Dan. Vid. Selsk. 32, no 12 (1960).
\bibitem{nilsson} S. G. Nilsson and O. Prior, Mat.
Fys. Medd. Dan. Vid. Selsk. 32, no 16 (1960).
\bibitem{coleman} A. J. Coleman, J. Math. Phys. 6, 1425 (1965).
\bibitem{wei2} L. Wei and C. C. Sun (2003), in preparation.
\end{thebibliography}
\end{document}